\documentclass[11pt,reqno]{amsart}
\usepackage{amssymb}

\usepackage[samelinetheorem,notitle]{maherart}
\usepackage{epigraph}
\usepackage{cancel}
\usepackage{mathrsfs}
\usepackage{bigints}
\usepackage{thmtools}
\usepackage{thm-restate}
\usepackage{standalone}
\usepackage{enumitem}
\usepackage{skak}
\usepackage{bbm}

\hypersetup{
	pdftitle =
	{Cloud Pricing},
	pdfauthor =
	{Dirk Bergemann  and Rahul Deb},
	pdfsubject=
	{Multi-Dimensional Screening, robust mechanism design, cloud pricing}
}
\newcommand{\mM}{\mathcal{M}}

\newcommand{\mF}{\mathcal{F}}

\newcommand{\Qbs}{Q^{b^{\star}}}
\newcommand{\Tbs}{T^{b^{\star}}}

\begin{document}
\title{Robust pricing for cloud computing}
\author[Bergemann]{Dirk Bergemann$^{\dag}$}
\address{$^{\dag}$Department of Economics, Yale University\\
\href{mailto:dirk.bergemann@yale.edu}{dirk.bergemann@yale.edu}}
\author[Deb]{Rahul Deb$^{\between}$ \\ \today}
\address{$^{\between}$Departments of Economics, Boston College and
University of Toronto\\
\href{mailto:rahul.deb@bc.edu}{rahul.deb@bc.edu}}

\begin{abstract}
We study the robust sequential screening problem of a monopolist seller of multiple cloud computing services facing a buyer who has private information about his demand distribution for these services. At the time of contracting, the buyer knows the distribution of his demand of various services and the seller simply knows the mean of the buyer's total demand. We show that a simple \textquotedblleft committed spend mechanism\textquotedblright\ is robustly optimal: it provides the seller with the highest profit
guarantee against all demand distributions that have the known total mean demand. This mechanism requires the buyer to commit to a minimum total usage and a corresponding base payment; the buyer can choose the individual quantities of each service and is free to consume additional units (over the committed total usage) at a fixed marginal price. This result provides theoretical support for prevalent cloud computing pricing practices while highlighting the robustness of simple pricing schemes in environments with
complex uncertainty. 

\medskip \noindent \textsc{Keywords:} Sequential Screening, Second-Degree Price Discrimination, Mechanism Design, Cloud Computing, Commitment Spend Contract.

\medskip

\noindent \textsc{JEL Classification:} D44, D82, D83.
\end{abstract}

\thanks{Dirk Bergemann gratefully acknowledges financial support from NSF SES 2049754 and ONR MURI. We thank Scott Shenker for productive conversations.}

\maketitle

\newpage

\section{Introduction}

\subsection{Motivation and Results}

Cloud computing has dramatically changed how individuals and businesses access computational resources, leading to fundamental theoretical questions about optimal pricing strategies in this market. Cloud computing providers offer several different services, perhaps most importantly: $\left( i\right)$ input/output, $\left( ii\right) $ compute, $\left( iii\right)$ memory/storage, and $\left(iv\right) $ software. A key challenge in designing cloud computing pricing mechanisms is that providers face significant uncertainty about customers' demands for these different services. Customers, themselves, may not know their exact demand at the time of contracting (as this will realize over a period of time) but they have knowledge about the \textit{distribution} of their demands of each service. This information is private and difficult for providers to verify or elicit. Providers must design pricing mechanisms that balance efficiency, profit, robustness and simplicity.

In practice, contracts for cloud computing frequently take the form of what are known as \textquotedblleft committed spend mechanisms.\textquotedblright\ Such a mechanism (or contract) commits the customer to spend a certain minimum amount $\underline{t}$ over a given time horizon. The contract is legally binding so the consumer cannot pay less than this minimum even if their realized demand for the cloud computing services is lower than expected. In return, the customer is allocated a \textit{total} quantity $\underline{q}$ of services and they are free to use any combination of individual quantities of different services that add up to less than this total. The contract additionally specifies a price $p$ and, if a customer's total usage of services $q$ exceeds $\underline{q}$, they are charged an additional amount $p(q-\underline{q})$ equal to their excess usage times the price.

\medskip

We model this setting as a robust multidimensional sequential screening problem. A seller (the provider of cloud computing services) has $K$ goods for sale and offers a contract. At the time of contracting, the buyer initially knows his distribution $F$ of demands for these different services. Upon contracting with the seller, the buyer learns his demand $(\theta_1,\dots,\theta_K)$ for the various services and chooses a quantity $(q_1,\dots,q_K)$ of each that maximizes his utility $\theta_1 u(q_1)+\cdots+\theta_K u(q_K)$ less the cost of this bundle of services.

Formally, the seller offers a mechanism---an allocation and transfer---that takes as its arguments a distribution and a realized demand type. The buyer first reports his type distribution and then his type, once it is realized. The seller simply knows the average total buyer demand $\lambda$ and chooses a robustly optimal mechanism that maximizes her worst-case profit evaluated against all possible buyer type (joint) distributions $F$ over $(\theta_1,\dots,\theta_K)$ that have this known average total demand (so $%
\mathbb{E}_F[\theta_1+\cdots+\theta_K]=\lambda$).

\medskip

The central result of this paper is to show that a committed spend mechanism---that we characterize explicitly---is robustly optimal. The main applied contribution of this result is that it provides an intuitive explanation for why committed spend mechanisms are widely used in cloud computing, highlighting their robustness and, as we argue below, efficiency. This result bridges the gap between theoretical mechanism design and real-world pricing practices, offering insights for practitioners and policymakers. Importantly, the optimal mechanism does not depend on, and therefore does not require the seller to solicit, the buyer's type distribution. In this multidimensional environment, the type distribution can be complex and hard to elicit in practice. Lastly, we show that the price $p$ for excess usage is simply the seller's cost. Consequently, conditional on the buyer having high-demand, the mechanism allocates efficiently.

\subsection{Related Literature}

This paper lies at the intersection of three literatures: sequential screening, multidimensional screening and robust mechanism design.

\medskip

Because the buyer in our model does not know their exact demand at the time of contracting, we are in a sequential screening environment. Following \citet{courty2000sequential}, the literature on sequential screening derives optimal contracts in a variety of different settings where buyers first learn their demand distribution for a single good and then their true demand is realized from this distribution in the second period. Because the buyer's private information realizes sequentially, the technical challenges associated with multidimensional screening can be skirted to a large extent. There are multiple goods in our model so all the technical challenges of multidimensional screening are present because the buyer learns his (multidimensional) type in the second period. We are nonetheless able to characterize the optimal mechanism because we employ a robust optimization
criterion. Consequently, we view our modeling choices to be an important contribution of this paper as, in our opinion, our model is both a natural fit for the pricing of cloud computing and nonetheless is tractable enough to characterize the (robustly) optimal mechanism in a multidimensional environment.

\medskip

The literature on multidimensional screening---in which the buyer knows his type at the moment of contracting---dates back till at least \citet{adams1976}. But, despite being a canonical problem, a general result characterizing the optimal mechanism (in a standard Bayesian environment) akin to \citet{myerson1981} (for the single good case) has proved elusive. \citet{schmalensee1984} and \citet{mcafee1989} showed that, for additive values and independent types, it is never optimal to sell items separately as the seller can do better by additionally offering the grand bundle at a discount (although this latter mechanism is also generally not optimal). \citet{bakos1999} showed that bundling a large number of independent products can improve profit due to the better predictability of consumers' valuations; the seller can extract almost all the surplus as the number of
goods gets large. \citet{armstrong2013} relaxed the assumption of additive values of two products and showed that even for potentially non-additive values, offering a bundle with a discounted price increases the profit whenever the demand for a single item is less elastic than the demand for the bundle.

The optimal mechanism for multiple products can take a very complex form.
Even for two goods and additive values that are independently drawn from the
same uniform distribution, \citet{pavlov2011} shows that the optimal
mechanism can involve randomization. When the values are drawn from a beta
distribution, \citet{DaDeTzEcma2017} show that the optimal mechanism can
feature an infinite menu of lotteries. In fact, \citet{hart2019} show that
profit maximizing mechanism that only involves pure bundling or selling
goods separately, might leave the seller with a negligible fraction of the
profit from the optimal mechanism. Thus, there is a disconnect between the
complex optimal mechanisms predicted by theory and the simple mechanisms we
observe in practice.

\medskip

This disconnect can be reduced by changing the optimization criterion from
Bayesian to robust. This was first observed by \citet{carroll2017} who
considers a seller who has information about the buyer's marginal value
distribution for each good but not the joint distribution. He shows that
selling each good separately is a robustly optimal mechanism when the seller
evaluates worst case profits against all joint distributions with the given
marginals. \citet{che2024} and \citet{deb2024} consider different robustness
criteria to that of Carroll and establish conditions under which it is
robustly optimal to sell bundles of goods. While bundling is commonly
observed in practice, the mechanisms in these papers require the seller to
randomize the bundle prices which is not something that real world sellers
do. Specifically, for our robustness criterion (that the seller knows the
mean of the total demand but nothing further), \citet{che2024} show that the
optimal mechanism involves pure bundling with a random bundle price.

\medskip

In contrast to these papers, we consider a multidimensional \textit{%
sequential screening} environment. Moreover, in our setting, there are 
\textit{multiple units} of each good for sale. That is, while the papers
mentioned in the previous paragraph are multidimensional versions of %
\citet{myerson1981}, our setting is a multdimensional, sequential screening
version of \citet{maskin1984}. In addition to realism, the advantage of the
sequential screening environment we study is that we show a committed spend
mechanism is robustly optimal that, in particular, does not require any
price randomization.

\section{Model}\label{sec:model}

We study the sequential screening problem of a single seller of $\mathcal{K}=\{1,\dots,K\}$ with $K\geq 1$ cloud computing services facing a single (representative) buyer. The seller offers a menu of contracts where each contract determines the price that the buyer would pay for using different quantities of the $K$ services. The buyer picks a contract from the menu, learns his computing needs and then determines his consumption of cloud computing services based on the prices of his chosen contract. In the remainder of this section, we describe this interaction formally.

\medskip

The buyer privately knows the (joint) distribution $F\in \Delta([0,1]^K)$ that determines his type $\theta=(\theta_1,\dots,\theta_K)\in [0,1]^K=:\Theta$ that, in turn, determines his demand for computing.\footnote{The restriction of the type space to $\Theta=[0,1]^K$ is simply a normalization to reduce notation. Instead of 1, we could have chosen the
upper bound arbitrarily and no result would change.} The value of a type $\theta$ for the quantity $q=(q_1\dots,q_K)\in \mathbb{R}^K_+$ of different computing services is given by  
\begin{equation*}
\sum_{i=1}^K \theta_i u(q_i),
\end{equation*}
where the utility function $u:\mathbb{R}_+\to \mathbb{R}_+$ is assumed to be strictly increasing and strictly concave with $u^{\prime }(0)=\infty$ and $\lim_{q_i\to \infty}u^{\prime }(q_i)=0$. We additionally assume that the buyer's preference is quasilinear in transfers.

The cost to the seller of providing the buyer with the services $q$ is 
\begin{equation*}
c\sum_{i=1}^K q_i
\end{equation*}
where $c\in\mathbb{R}_{++}$.

The buyer's type distribution $F$ lies in a set $\mathcal{F}\subseteq \Delta
([0,1]^{K})$ where 
\begin{equation*}
\mathcal{F}:=\left\{ F\in \Delta ([0,1]^{K})\;\bigg|\;\int_{\Theta }(\theta
_{1}+\cdots +\theta _{K})dF(\theta )=\lambda \right\} .
\end{equation*}%
In words, the distribution from which the buyer's type is drawn can be any distribution such that the expectation of the sum of the $K$ dimensions is $\lambda \in (0,K)$.

\medskip

The timing of the interaction between the buyer and seller is the following: 

\begin{enumerate}
\item The seller offers the buyer a mechanism $(Q,T)$ where $Q$ is the allocation and $T$ is the transfer. We formally define a mechanism below. 

\item The buyer is privately informed about her distribution $F\in\mathcal{F}$, reports $F^{\prime }\in\mathcal{F}$ to the mechanism. 

\item The buyer's value $\theta$ is drawn from distribution $F$. The buyer reports value $\theta^{\prime }\in\Theta$. 

\item The buyer and seller receive their respective payoffs from the
mechanism. 
\end{enumerate}

\medskip

The revelation principle applies so it is without loss to assume that, in step 1, the seller offers the buyer a (direct) mechanism  
\begin{equation*}
Q:\mathcal{F}\times \Theta \to \mathbb{R}^K_+\;\; \text{ and }\;\; T:\mathcal{F}\times \Theta \to \mathbb{R},
\end{equation*}
that satisfies incentive compatibility and individual rationality. The function $Q(F^{\prime },\theta^{\prime })$ determines the quantity of each service that the buyer receives when she reports $F^{\prime }$ in step 2 and $\theta^{\prime }$ in step 3. $T(F^{\prime },\theta^{\prime })$ is the transfer from the buyer to the seller.

\medskip

Incentive compatibility (or IC for short) requires that 
\begin{equation*}
\sum_{i=1}^{K}\theta _{i}u(Q_{i}(F,\theta ))-T(F,\theta )\geq \sum_{i=1}^{K}\theta _{i}u(Q_{i}(F,\theta ^{\prime }))-T(F,\theta ^{\prime }) \text{ for all }\theta ,\theta ^{\prime }\in \Theta \text{ and all }F\in \mathcal{F}
\end{equation*}%
and 
\begin{equation*}
\mathbb{E}_{F}\left[ \sum_{i=1}^{K}\theta _{i}u(Q_{i}(F,\theta ))-T(F,\theta)\right] \geq \mathbb{E}_{F}\left[ \sum_{i=1}^{K}\theta_{i}u(Q_{i}(F^{\prime },\theta ))-T(F^{\prime },\theta )\right] \text{ for all }F,F^{\prime }\in \mathcal{F}.
\end{equation*}%
The incentive compatibility (IC) constraint can be written this way (as opposed to in terms of joint misreports of $F$ and $\theta $) because, as in standard sequential screening, once the buyer's value $\theta $ is realized, the distribution $F$ no longer affects their utility.

\medskip

Individual rationality (or IR for short) requires that  
\begin{equation*}
\mathbb{E}_F\left[\sum_{i=1}^K \theta_i u(Q_i(F,\theta))-T(F,\theta)\right] \geq 0 \text{ for all }F\in \mathcal{F}.
\end{equation*}
We denote the set of all IC and IR mechanisms as $\mathcal{M}$.

\medskip

The seller solves the following robust mechanism design problem  
\begin{equation}  \label{eq:seller_prob}
(Q^{\star},T^{\star})\in\argmax_{(Q,T)\in\mathcal{M}} \inf_{F\in\mathcal{F}}
\left\{\mathbb{E}_{F}\left[T(F,\theta) - c\sum_{i=1}^K Q_i(F,\theta)\right]%
\right\}.
\end{equation}
In words, the seller chooses a mechanism that maximizes her payoff again the worst-possible type distribution in the set $\mathcal{F}$ that the buyer might have. We refer to a solution $(Q^{\star},T^{\star})$ as a \textit{robustly optimal mechanism}.

\medskip

Before we present our main result in the next section, it is worth making a few remarks. First, note that the statement \eqref{eq:seller_prob} of the seller's problem implicitly assumes that a robustly optimal mechanism exists. We will show that this is indeed the case. 

Second, observe that the allocation $Q$ is restricted to be deterministic; there is obviously no benefit in randomizing the transfer since this is enters linearly in the
payoff of both the seller and the buyer. As we argue next, the restriction to deterministic allocations is without loss.

\begin{lemma}
	Let $$\mM^r:=\{(Q,T)\;|\; Q:\mathcal{F}\times \Theta \to \Delta(\mathbb{R}^K_+),\; T:\mathcal{F}\times \Theta \to \Delta(\mathbb{R}) \text{ and } (Q,T) \text{ is IC and IR } \}$$ be the set of incentive compatible and individually rational random mechanisms. Then,
	$$\max_{(Q,T)\in\mathcal{M}^r} \inf_{F\in\mathcal{F}}
	\left\{\mathbb{E}_{F}\left[T(F,\theta) - c\sum_{i=1}^K Q_i(F,\theta)\right]%
	\right\}=\max_{(Q,T)\in\mathcal{M}} \inf_{F\in\mathcal{F}}
	\left\{\mathbb{E}_{F}\left[T(F,\theta) - c\sum_{i=1}^K Q_i(F,\theta)\right]%
	\right\}$$
	or, in words, that the seller does not achieve a higher profit guarantee by employing a random mechanism.
\end{lemma}

\begin{proof}	
	The proof of the lemma employs a simple argument: we take any mechanism that involves randomization and then construct an alternate deterministic mechanism that yields the
	seller a higher payoff.
	
	Consider an IC and IR mechanism $(Q,T)$ in which both the allocation $Q:\mathcal{F}\times \Theta \to \Delta(\mathbb{R}^K_+)$ and transfer $T:\mathcal{F}\times \Theta \to \Delta(\mathbb{R})$ can be random.
	
	Construct a deterministic mechanism $(Q^{\prime },T^{\prime })$	such that  
	\begin{equation*}
		u(Q^{\prime }_i(F,\theta))=\mathbb{E}_{Q_i(F,\theta)} [u(q_i)] \text{ and }
		T^{\prime }(F,\theta)=\mathbb{E}_{T(F,\theta)}[t] \;\; \text{ for all } i\in%
		\mathcal{K},\; F\in\mathcal{F} \text{ and } \theta\in\Theta,
	\end{equation*}
	where the first expectation is taken over $q_i$ with respect to the	(marginal) allocation distribution $Q_i(F,\theta)$ and the second expectation is taken over $t$ with respect to the transfer distribution $T(F,\theta)$.
	
	\medskip
	
	By construction, we have that  
	\begin{equation*}
		\sum_{i=1}^K \theta_i u(Q^{\prime }_i(F,\theta^{\prime })) - T^{\prime}(F,\theta^{\prime }) = \sum_{i=1}^K \theta_i \mathbb{E}_{Q_i(F,\theta^{\prime })} [u(q_i)] - \mathbb{E}_{T(F,\theta^{\prime })}[t],
	\end{equation*}
	where the left and right sides are the utilities of a buyer who reported distribution $F\in\mathcal{F}$ in step 2, had type $\theta\in\Theta$ realize but reported type $\theta^{\prime }\in\Theta$ in step 3 from mechanisms $(Q^{\prime },T^{\prime })$ and $(Q,T)$ respectively. Consequently, the fact that $(Q,T)$ is IC and IR implies that the deterministic mechanism $(Q^{\prime },T^{\prime })$ is also IC and IR.
	
	Lastly observe that, since $u$ is concave, $u(Q^{\prime }_i(F,\theta))=\mathbb{E}_{Q_i(F,\theta)} [u(q_i)]$ implies that the certainty equivalent $Q^{\prime }_i(F,\theta)\leq \mathbb{E}_{Q_i(F,\theta)} [q_i]$ for all $i\in	\mathcal{K}$, $F\in\mathcal{F}$ and $\theta\in\Theta$. Consequently, the seller's expected cost  
	\begin{equation*}
		\int_{\Theta} c \sum_{i=1}^K Q^{\prime }_i(F,\theta) dF(\theta) \leq
		\int_{\Theta} c \sum_{i=1}^K \mathbb{E}_{Q_i(F,\theta)} [q_i] dF(\theta)
	\end{equation*}
	is lower for mechanism $(Q^{\prime },T^{\prime })$ for all distributions $F\in\mathcal{F}$.
	
	Since, the expected transfers from both $(Q^{\prime },T^{\prime })$ and $(Q,T)$ are same by construction, this implies that the seller's worst case	profit (taken over $F\in\mathcal{F}$) from $(Q^{\prime },T^{\prime })$ is weakly greater than that from $(Q,T)$. This shows that the seller has no
	incentive to choose a mechanism with a randomized allocation and transfer.
\end{proof}

In words, because the buyer's utility is strictly concave in the quantity of each service, the seller has no incentive to offer a random allocation rule $Q:%
\mathcal{F}\times \Theta \to \Delta(\mathbb{R}^K_+)$ since she could instead offer certainty equivalents at a lower cost without changing buyer incentives. We chose not to allow random allocations while defining the model above as this needlessly complicates the presentation.

\section{The optimality of committed spend mechanisms}\label{sec:main_result}

Before we present our main result, we need one additional definition.

\medskip

A \textit{committed spend mechanism} $(Q^{b},T^{b})\in \mathcal{M}$ is
defined by three values $\underline{q}\in \mathbb{R}_{+}$, $\underline{t}\in 
\mathbb{R}_{+}$ and $p\in \mathbb{R}_{++}$ that determine the allocation and
transfer as follows: 
\begin{align*}
& Q^{b}(F,\theta )\in \argmax_{(q_{1},\dots ,q_{K})}\left\{
\sum_{i=1}^{K}\theta _{i}u(q_{i})-\underline{t}-p\left( \sum_{i=1}^{K}q_{i}-%
\underline{q}\right) \right\} \text{ such that }\sum_{i=1}^{K}q_{i}\geq 
\underline{q}, \\
& T^{b}(F,\theta )=\underline{t}+p\left( \sum_{i=1}^{K}Q_{i}^{b}(F,\theta )-%
\underline{q}\right) .
\end{align*}%
In words, a committed spend mechanism requires the buyer to commit to a
consuming a minimal total (summed across services) quantity $\underline{q}$
and making a minimal payment $\underline{t}$. The buyer can choose to
consume a higher total quantity and pay for the additional quantity consumed
at a unit price $p$. The buyer is free to pick how she allocates the total
quantity across different services. Note that, our assumptions on $u$ imply
that $Q^{b}$ is well defined as there is always a solution to the
maximization problem that defines the quantity consumed. Indeed, the
solution is unique for all $\theta \neq 0$.

\medskip

As we discussed earlier, cloud computing is typically priced using a
committed spend mechanism. Our main result shows that there is a robustly
optimal mechanism of this form!

\begin{theorem}\label{thm:main_result}
		The committed spend mechanism $(\Qbs,\Tbs)$ defined by
		$$\underline{q}^{\star}=Ku'^{-1}\left(\frac{cK}{\lambda}\right),\; \underline{t}^{\star}=\lambda u\left(\frac{\underline{q}^{\star}}{K}\right) \text{ and } p^{\star}=c$$
		is a robustly optimal mechanism; that is, it is a solution to the principal's problem \eqref{eq:seller_prob}.
	\end{theorem}

This result states that there is a robust optimal mechanism in which the
buyer commits to using at least a total $\underline{q}^{\star}$ of the
various cloud computing services and paying at least $\underline{t}^{\star}$%
. The buyer has the option of consuming a higher quantity, should he need
too, and additional units are priced by the seller at her cost $c$.

Observe a few nice properties of this robustly optimal mechanism. Such
mechanisms are observed in practice and do not have the unrealistic feature
of requiring the buyer to report his type distribution $F$. Moreover, the
mechanism is such that, whenever the buyer consumes a higher total quantity
than $\underline{q}^{\star }$, his chosen quantity is the socially efficient
level since it maximizes $\sum_{i=1}^{K}\theta
_{i}u(q_{i})-c\sum_{i=1}^{K}q_{i}$! As the proof of \cref{thm:main_result}
in the next section shows, there are multiple robustly optimal mechanisms
and this feature makes the committed spend mechanism particularly appealing.

\section{The Proof of \cref{thm:main_result}}\label{sec:proof}

The result is proved in several steps. We first establish a robustly optimal
mechanism for the special case of a single service ($K=1$). We then use that
to derive a robustly optimal mechanism for the case of multiple services ($%
K>1$). This mechanism is not a committed spend mechanism. In the final step,
we show that there is also a committed spend mechanism that is robustly
optimal and strictly more efficient.

\medskip

The first lemma derives a robustly optimal mechanism for a single service
(that is, $K=1$).  
\begin{lemma}\label{lem:one_good}
		Suppose $K=1$. There is a robust optimal mechanism $Q^*(F,\theta)=q^*$, $T^*(F,\theta)=\lambda u(q^*)$ for all $F\in\mF$ and $\theta\in [0,1]$ where $q^*=u'^{-1}\left(\frac{c}{\lambda}\right)$.
	\end{lemma}

\begin{proof}
Let $\delta_{\lambda}\in \mathcal{F}$ denote the Dirac distribution that
assigns probability one to the mean $\lambda$.

\medskip

Take any IC and IR mechanism $(Q,T)\in \mathcal{M}$. Observe that because $%
(Q,T)$ is IR, we have  
\begin{equation*}
\mathbb{E}_{\delta_{\lambda}}\left[\theta
u(Q(\delta_{\lambda},\theta))-T(\delta_{\lambda},\theta)\right] \geq 0 \iff
T(\delta_{\lambda},\lambda)\leq \lambda u(Q(\delta_{\lambda},\lambda)).
\end{equation*}

\medskip

Now define a new mechanism $(\hat{Q},\hat{T})$ such that  
\begin{equation*}
\hat{Q}(F,\theta)=Q(\delta_{\lambda},\lambda),\; \hat{T}(F,\theta)=\lambda
u(Q(\delta_{\lambda},\lambda)) \text{ for all } F\in \mathcal{F} \text{ and }
\theta\in [0,1].
\end{equation*}
In words, the mechanism $(\hat{Q},\hat{T})$ is constant in both $F$ and $%
\theta$ with the allocation and transfer being $Q(\delta_{\lambda},\lambda)$
and $\lambda u(Q(\delta_{\lambda},\lambda))$ respectively.

\medskip

$(\hat{Q},\hat{T})$ is clearly IC since it is a constant mechanism and so
misreporting $F$ and $\theta$ makes no difference to either the allocation
or the transfer. Also,  
\begin{equation*}
\mathbb{E}_F\left[ \theta u(\hat{Q}(F,\theta))-\hat{T}(\theta)\right] = 
\mathbb{E}_F[\theta] u(Q(\delta_{\lambda},\lambda))-\lambda
u(Q(\delta_{\lambda},\lambda)) = 0 \;\; \text{ for all }F\in \mathcal{F}
\end{equation*}
because $\mathbb{E}_F[\theta]=\lambda$. Consequently, $(\hat{Q},\hat{T})$ is
also IR and so $(\hat{Q},\hat{T})\in\mathcal{M}$.

\medskip

Therefore, for any $(Q,T)\in\mathcal{M}$, we have  
\begin{align*}
\inf_{F\in \mathcal{F}} \mathbb{E}_F\left[T(F,\theta) - c Q(F,\theta)\right]
& \leq \mathbb{E}_{\delta_{\lambda}} \left[T(\delta_{\lambda},\theta) - c
Q(\delta_{\lambda},\theta)\right] \\
& = T(\delta_{\lambda},\lambda) - c Q(\delta_{\lambda},\lambda) \\
& \leq \lambda u(Q(\delta_{\lambda},\lambda)) - c Q(\delta_{\lambda},\lambda)
\\
& = \inf_{F\in \mathcal{F}} \mathbb{E}_F\left[ \hat{T}(F,\theta) - c \hat{Q}%
(F,\theta)\right].
\end{align*}
In words, relative to any mechanism $(Q,T)\in\mathcal{M}$, the principal
gets a higher profit guarantee offering the mechanism $(\hat{Q},\hat{T})\in%
\mathcal{M}$ instead. Thus, there is a robustly optimal mechanism that takes
the form $(\hat{Q},\hat{T})$ (were a robustly optimal mechanism to exist).

\medskip

Let,  
\begin{equation*}
q^*=\argmax_q \{\lambda u(q) - c q\}
\end{equation*}
and so, from the first-order condition, we have  
\begin{equation*}
q^*=u^{\prime -1}\left(\frac{c}{\lambda}\right).
\end{equation*}
In words, this states that, if we were to consider mechanisms that always
offer the constant quantity $q$ along with a transfer $\lambda u(q)$, the
highest possible profit is obtained by choosing the quantity $q^*=u^{\prime
-1}\left(\frac{c}{\lambda}\right).$ Combining this observation with the
above argument, we have shown that there is a robustly optimal mechanism  
\begin{equation*}
Q^*(F,\theta)=q^*,\; T^*(F,\theta)=\lambda u(q^*) \text{ for all } F\in%
\mathcal{F} \text{ and } \theta\in [0,1]
\end{equation*}
which completes the proof of the lemma. 
\end{proof}

\medskip

We now move to the case where there are multiple services offered by the
seller (so $K>1$). We begin with some notation. Let  
\begin{equation*}
\Theta^d:=\{(\tilde{\theta},\dots,\tilde{\theta})\;|\;\tilde{\theta}\in[0,1]%
\}\subset \Theta
\end{equation*}
denote the subset of types that lie on the diagonal. We use $\mathcal{F}%
^{d}\subset \mathcal{F}$ to denote the subset of (perfectly correlated)
distributions that are defined on the set $\Theta^d$, that is, $%
F^d(\Theta^d)=1$ for all $F^d\in\mathcal{F}^d$.\footnote{%
In a slight abuse of notation, we use $F$ to refer both to a cumulative
distribution and a measure. The meaning should be clear from the argument
(type or set of types respectively).}

As an intermediate step, we now derive a robustly optimal mechanism for the
case where the buyer's type distribution $F^d$ lies in $\mathcal{F}^d$ and
consequently, the buyer's type lies on the diagonal $\Theta^d$. Formally, we
now solve  
\begin{equation}  \label{eq:seller_prob_PC}
(Q^{d^{\star}},T^{d^{\star}})\in\argmax_{(Q,T)\in\mathcal{M}} \inf_{F^d\in%
\mathcal{F}^d} \left\{\mathbb{E}_{F^d}\left[T(F^d,\theta) - c\sum_{i=1}^K
Q_i(F^d,\theta)\right]\right\},
\end{equation}
the solution to which is described in the next lemma.

\medskip

\begin{lemma}\label{lem:diagonal}
		There is a solution to the problem \eqref{eq:seller_prob_PC} given by $Q_i^{d^{\star}}(F,\theta)=q^{d^{\star}},\;\; T^{d^{\star}}(F,\theta)=\lambda u(q^{d^{\star}})$ for all $F\in\mF$ and $\tilde{\theta}\in \Theta$ where $q^{d^{\star}}=u'^{-1}\left(\frac{cK}{\lambda}\right)$ $q^*$.
	\end{lemma}

\begin{proof}
Take any IC and IR mechanism $(Q,T)\in \mathcal{M}$ and any distribution $%
F^d\in\mathcal{F}^d$. Define an alternate mechanism $(Q^d,T^d)$ where the
allocation for types along the diagonal is given by the solution to  
\begin{equation*}
u(Q^d_i(F^d,(\tilde{\theta},\dots,\tilde{\theta})))=\frac{1}{K}\sum_{j=1}^K
u(Q_j(F^d,(\tilde{\theta},\dots,\tilde{\theta}))) \;\; \text{ for all } \;\; 
\tilde{\theta}\in[0,1], \;i\in\mathcal{K}
\end{equation*}
and the allocation  
\begin{equation*}
Q^d(F^d,\theta)=Q^d\left(F^d,\left(\frac{\theta_1+\cdots+\theta_K}{K},\dots,%
\frac{\theta_1+\dots+\theta_K}{K}\right)\right)
\end{equation*}
for all other types $\theta\in\Theta\backslash\Theta^d$ is defined to be the
allocation corresponding to the diagonal type $(\tilde{\theta},\dots,\tilde{%
\theta})$ where $K\tilde{\theta}=\theta_1+\dots+\theta_K$.

\medskip

The transfers $T^d$ are the same  
\begin{equation*}
T^d(F^d,\theta^d)=T(F^d,\theta^d)
\end{equation*}
for types $\theta^d\in\Theta^d$ that lie on the diagonal. For all other
types $\theta\in\Theta\backslash\Theta^d$, the transfer  
\begin{equation*}
T^d(F^d,\theta)=T\left(F^d,\left(\frac{\theta_1+\dots+\theta_K}{K},\dots,%
\frac{\theta_1+\dots+\theta_K}{K}\right)\right)
\end{equation*}
is the same as the diagonal type $(\tilde{\theta},\dots,\tilde{\theta})$
where $K\tilde{\theta}=\theta_1+\dots+\theta_K$.

\medskip

For any distribution $F\in\mathcal{F} \backslash \mathcal{F}^d$ that is not
in the set $\mathcal{F}^d$, let $G^d_F\in\mathcal{F}^d$ be the distribution
such that  
\begin{equation*}
G^d_F(\tilde{\theta},\dots,\tilde{\theta})=F(\{(\theta\in\Theta\;|\;
\theta_1+\cdots+\theta_K\leq K\tilde{\theta})\}) \;\; \text{ for all }\tilde{%
\theta}\in[0,1].
\end{equation*}
In words, $G^d_F\in\mathcal{F}^d$ is the distribution supported on the
diagonal $\Theta^d$ that induces the same distribution on the sum $%
\theta_1+\cdots+\theta_K$ as $F$. Note that the distribution $G^d_F\in%
\mathcal{F}^d$ always exists for all $F\in\mathcal{F}$.

\medskip

Now, for all $F\in\mathcal{F} \backslash \mathcal{F}^d$, we define  
\begin{equation*}
Q^d(F,\theta)=Q^d(G^d_F,\theta) \text{ and } T^d(F,\theta)=T^d(G^d_F,\theta)
\end{equation*}
where $Q^d(G^d_F,\theta)$ and $T^d(G^d_F,\theta)$ are defined above.

\medskip

Observe that, for any reported $F^d\in\mathcal{F}^d$, if type $(\tilde{\theta%
},\dots,\tilde{\theta})$ reports as type $(\tilde{\theta}^{\prime },\dots,%
\tilde{\theta}^{\prime })$, their payoff in the mechanism $(Q^d,T^d)$  
\begin{equation*}
\sum_{i=1}^K \tilde{\theta}u(Q_i^d(F^d,(\tilde{\theta}^{\prime },\dots,%
\tilde{\theta}^{\prime d}(F^d,(\tilde{\theta}^{\prime },\dots,\tilde{\theta}%
^{\prime }))=\tilde{\theta}\sum_{i=1}^K u(Q_i(F^d,(\tilde{\theta}^{\prime
},\dots,\tilde{\theta}^{\prime d},(\tilde{\theta}^{\prime },\dots,\tilde{%
\theta}^{\prime }))
\end{equation*}
is the same as that from the original mechanism $(Q,T)$. Moreover, for any
type $\theta\in\Theta\backslash\Theta^d$ that is not on the diagonal, the
allocation $Q^d$ and transfer $T^d$ is the same as the type $\left(\frac{%
\theta_1+\dots+\theta_K}{K},\dots,\frac{\theta_1+\dots+\theta_K}{K}\right)$.
Moreover, for all types, the same quantity of each service is allocated.
Consequently, the payoff that type $\theta$ gets from misreporting as any
other type $\theta^{\prime }\in\Theta$ is the same as the payoff that type $%
\left(\frac{\theta_1+\dots+\theta_K}{K},\dots,\frac{\theta_1+\dots+\theta_K}{%
K}\right)$ receives by misreporting as $\left(\frac{\theta^{\prime
}_1+\dots+\theta^{\prime }_K}{K},\dots,\frac{\theta^{\prime
}_1+\dots+\theta^{\prime }_K}{K}\right)$. Therefore, because the original
mechanism $(Q,T)$ is IC, the buyer has no incentive to misreport their type $%
\theta$ (in step 3) after reporting a distribution $F^d\in\mathcal{F}^d$ in
step 2.

Since we defined $Q^d(F,\theta)=Q^d(G^d_F,\theta)$ and $T^d(F,%
\theta)=T^d(G^d_F,\theta)$, the above argument also implies that the buyer
has no incentive to misreport their type $\theta$ if they report a
distribution $F\in\mathcal{F} \backslash \mathcal{F}^d$ in step 2.

\medskip

Observe that  
\begin{equation*}
\mathbb{E}_{F^d}\left[\sum_{i=1}^K \theta_i u(Q^d_i(\tilde{F}^d,\theta))-T^d(%
\tilde{F}^d,\theta)\right] = \mathbb{E}_{F^d}\left[\sum_{i=1}^K \theta_i
u(Q_i(\tilde{F}^d,\theta))-T(\tilde{F}^d,\theta)\right] \text{ for all } F^d,%
\tilde{F}^d\in \mathcal{F}^d
\end{equation*}
because we have already argued that the type $\theta$ is truthfully reported
in step 3. This implies that since, $(Q,T)$ is IC, the buyer whose type
distribution is $F^d,\tilde{F}^d\in \mathcal{F}^d$ has no incentive to
misreport as $F^d,\tilde{F}^d\in \mathcal{F}^d$. Moreover, since $(Q,T)$ is
IR, the buyer's expected utility from reporting both $F^d$ and $\theta$
truthfully is at least zero.

\medskip

For any $F\in\mathcal{F} \backslash \mathcal{F}^d$ and $F^{\prime }\in 
\mathcal{F}$, observe that  
\begin{equation*}
\mathbb{E}_{F}\left[\sum_{i=1}^K \theta_i u(Q^d_i(F^{\prime d}(F^{\prime
},\theta)\right] =\mathbb{E}_{G^d_F}\left[\sum_{i=1}^K \theta_i
u(Q^d_i(G^d_{F^{\prime }},\theta))-T^d(G^d_{F^{\prime }},\theta)\right].
\end{equation*}
In words, the expected utility that a buyer with type distribution $F$
receives from the mechanism $(Q^d,T^d)$ when she misreports as type
distribution $F^{\prime }$ is the same as the utility that a buyer with type
distribution $G^d_F$ gets from misreporting as $G^d_{F^{\prime }}$. We have
already argued above that such a misreport does not benefit the buyer.

Thus, taken together, we have shown that the mechanism $(Q^d,T^d)$ is IC, IR
and provides the same expected utility as $(Q,T)$ whenever the buyer's type
distribution is $F^d\in\mathcal{F}^d$.

\medskip

Now, note that the concavity of $u$ implies that  
\begin{equation}  \label{eq:concavity}
u(Q_i^d(\tilde{\theta},\dots,\tilde{\theta}))\leq u\left( \frac{1}{K}%
\sum_{j=1}^K Q_j(\tilde{\theta},\dots,\tilde{\theta})\right) \implies K
Q_i^d(\tilde{\theta},\dots,\tilde{\theta})\leq \sum_{j=1}^K Q_j(\tilde{\theta%
},\dots,\tilde{\theta})
\end{equation}
where the second inequality follows from the fact that $u$ is strictly
increasing.

Thus, for any $F^d\in \mathcal{F}^d$, the principal's payoff from $(\hat{Q}%
^d,T^d)$  
\begin{equation*}
\mathbb{E}_{F^d}\left[T^d(\tilde{\theta},\dots,\tilde{\theta})-c\sum_{i=1}^K
Q^d_i(\tilde{\theta},\dots,\tilde{\theta})\right] \geq \mathbb{E}_{F^d}\left[%
T(\tilde{\theta},\dots,\tilde{\theta})-c\sum_{i=1}^K Q_i(\tilde{\theta}%
,\dots,\tilde{\theta})\right]
\end{equation*}
is higher than the original mechanism $(Q,T)$ because the transfer is the
same for all types in the support of $F^d$ and, the total cost is lower (due
of inequality \ref{eq:concavity}).

Thus, in order to solve the problem \ref{eq:seller_prob_PC}, it is without
loss for the seller to consider mechanisms where the quantity $%
Q_1(\cdot)=\cdots=Q_K(\cdot)$ is the same for all goods. This, in turn,
implies that the seller's problem collapses to that of a single good. The
characterization of the solution to the seller's problem %
\eqref{eq:seller_prob_PC} then follows from \cref{lem:one_good}. 
\end{proof}

We use \cref{lem:diagonal} to characterize a robustly optimal mechanism for
the seller. This is not a committed spend mechanism but we will show that a
committed spend mechanism is also robustly optimal and strictly more
efficient.

\begin{lemma}\label{lem:robut_opt}
		There is a robust optimal mechanism $Q^*_i(F,\theta)=q^*$, $T^*(F,\theta)=\lambda u(q^*)$ for all $F\in\mF$ and all $\theta\in \Theta$ where $q^*=u'^{-1}\left(\frac{cK}{\lambda}\right)$
	\end{lemma}

\begin{proof}
Observe that  
\begin{align*}
\max_{(Q,T)\in \mathcal{M}} \inf_{F\in \mathcal{F}} \mathbb{E}_F\left[%
T(F,\theta) - c\sum_{i=1}^K Q_i(F,\theta)\right] & \leq \max_{(Q,T)\in 
\mathcal{M}} \inf_{F^d\in \mathcal{F}^d} \mathbb{E}_{F^d}\left[T(\theta) -
c\sum_{i=1}^K Q_i(\theta)\right] \\
& = \lambda u(q^*)- cKq^* \\
& = \inf_{F\in \mathcal{F}} \mathbb{E}_F\left[T^*(F,\theta) - c\sum_{i=1}^K
Q^*_i(F,\theta)\right],
\end{align*}
where the inequality follows from the fact that the infimum is being taken
over a smaller set, the first equality follows from \cref{lem:diagonal} and
the last inequality follows from the definition of $(Q^*,T^*)$ in the
statement of the lemma. The robust optimality of $(Q^*,T^*)$ follows. 
\end{proof}

\cref{thm:main_result} is an immediate consequence of \cref{lem:robut_opt}
and we complete the proof below.

\begin{proof}[Proof of \cref{thm:main_result}]
Consider the committed spend mechanism defined in the statement of the
theorem and recall that a committed spend mechanism does not depend on the
reported type distribution $F\in \mathcal{F}$. We now compute the seller's
profit from each type $\theta \in \Theta $ and show that it is the same as
the profit the seller obtains from the mechanism $(Q^{\ast },T^{\ast })$.

For each $F\in\mathcal{F}$ and $\theta\in\Theta$, observe that the seller's
payoff is  
\begin{align*}
T^{b^{\star}}(F,\theta) - c \sum_{i=1}^K Q^{b^{\star}}_i(F,\theta) & =
\lambda u\left(\frac{\underline{q}^{\star}}{K}\right) + c\left(\sum_{i=1}^K
Q^{b^{\star}}_i(F,\theta)-\underline{q}^{\star}\right) - c \underline{q}%
^{\star} - c\left(\sum_{i=1}^K Q^{b^{\star}}_i(F,\theta)-\underline{q}%
^{\star}\right) \\
& = \lambda u\left(\frac{\underline{q}^{\star}}{K}\right) - c \underline{q}%
^{\star} \\
& = \lambda u(q^*)- cKq^* \\
& = T^*(F,\theta) - c\sum_{i=1}^K Q^*_i(F,\theta).
\end{align*}

The committed spend mechanism is IC by definition. Note that for all $F\in 
\mathcal{F}$ and all $\theta \in \Theta $, the buyer's payoff is at least as
high as his payoff from $(Q^{\ast },T^{\ast })$. Consequently, $(Q^{b^{\star
}},T^{b^{\star }})$ is also IR (that is, $(Q^{b^{\star }},T^{b^{\star }})\in 
\mathcal{M}$) and its robust optimality follows from the robust optimality
of $(Q^{\ast },T^{\ast })$. This completes the proof of the theorem. 
\end{proof}

\section{Conclusion}

This paper has examined the problem of pricing in cloud computing through
the lens of robust mechanism design. By formulating a sequential screening
model, we have characterized the optimal mechanism for a seller facing a
buyer with private information about their type distribution. Our main
result demonstrates that a committed spend mechanism---requiring a minimum
commitment and pricing additional usage at cost---is robustly optimal.

\medskip

Three key insights emerge from our analysis. First, the ability to achieve
optimal profits with mechanisms that do not require buyers to report their
demand distributions indicates that simpler pricing schemes may be
preferable to more complex ones. Second, the flexibility afforded to buyers
in allocating their usage across services aligns with practical needs while
preserving optimality. Finally, the optimality of setting marginal prices
equal to marginal costs for usage above the minimum commitment suggests that
efficiency considerations should guide pricing of incremental consumption.

The findings have significant implications for the design of cloud computing
contracts. committed spend mechanisms are not only theoretically optimal but
also align closely with practical pricing schemes observed in the industry.
Their robustness to information asymmetry and efficiency in resource
allocation make them particularly appealing in dynamic and uncertain
environments. Moreover, the simplicity of these mechanisms enhances their
feasibility for implementation, reducing the informational burden on both
buyers and sellers.

Future research could explore extensions of this framework to multi-seller
markets, dynamic pricing scenarios, and environments with additional sources
of uncertainty. By doing so, we can further deepen our understanding of
optimal pricing strategies in cloud computing and related industries.

\medskip

To conclude, this paper demonstrates that simple committed spend mechanisms
are optimal for pricing cloud computing services under demand uncertainty.
Our results provide theoretical support for common industry practices while
highlighting how elementary pricing structures can achieve optimal outcomes
in complex environments. The robustness of committed spend
mechanisms---their ability to achieve optimal profits without detailed
knowledge of demand patterns---may help explain their widespread adoption in
cloud computing markets.

\newpage

\bibliographystyle{econometrica}
\bibliography{CloudPricingRefs}

\end{document}